\documentclass[%
 reprint,
 amsmath,amssymb,
 aps,
]{revtex4-2}

\usepackage[ruled,vlined]{algorithm2e}
\usepackage{enumerate}
\usepackage{chngcntr}
\usepackage{graphicx}
\usepackage{dcolumn}
\usepackage{tikz}
\usetikzlibrary{positioning} 
\usetikzlibrary{matrix}
\usepackage{xcolor}
\usepackage{soul}
\usepackage{bm}
\usepackage{hyperref}
\usepackage[mathlines]{lineno}
\usepackage{etoolbox}

\makeatletter
\patchcmd{\@algocf@start}
  {-1.5em}
  {0pt}
  {}{}
\makeatother

\begin{document}

\title{Dynamical importance and network perturbations}

\author{Ethan Young$^{1,}$}
\email{young.j.ethan@gmail.com}
\altaffiliation[now at the ]{Department of Applied Mathematics, University of Washington}
\email{ethanjy@uw.edu}
\author{Mason A. Porter$^{1, 2, 3, }$}
\email{mason@math.ucla.edu}
\affiliation{
    $^1$Department of Mathematics, University of California,\\
    Los Angeles, CA, 90095, United States of America\\
    $^2$Department of Sociology, University of California,\\
    Los Angeles, CA, 90095, United States of America\\
    $^3$Santa Fe Institute, Santa Fe, NM, 87501, United States of America
}

\date{\today}

\begin{abstract}
The leading eigenvalue $\lambda$ of the adjacency matrix of a graph exerts much influence on the behavior of dynamical processes on that graph. It is thus relevant to relate notions of the importance (specifically, centrality measures) of network structures to $\lambda$ and its associated eigenvector. We study a previously derived measure of edge importance known as ``dynamical importance'', which estimates how much $\lambda$ changes when one removes an edge from a graph or adds an edge to it. We examine the accuracy of this estimate for different network structures and compare it to the true change in $\lambda$ after an edge removal or edge addition. We then derive a first-order approximation of the change in the leading eigenvector. We also consider the effects of edge additions on Kuramoto dynamics on networks, and we express the Kuramoto order parameter in terms of dynamical importance. Through our analysis and computational experiments, we find that studying dynamical importance can improve understanding of the relationship between network perturbations and dynamical processes on networks.
\end{abstract}


\maketitle


\section{Introduction}

The study of dynamical processes on graphs and other networks is important for many applications, which range from disease spread in populations to the collective behavior of neurons in biological neural networks \cite{porter2016,newman2018}. A major research direction is the examination of how dynamical processes are affected by network architecture. Consider a network in the form of a graph, in which nodes are connected pairwise (i.e., dyadically) by edges. There is an intimate relationship between a graph's structure and the spectral properties (i.e., the eigenvalues and associated eigenvectors) of its adjacency matrix $A$ (and of other matrices, such as Laplacian matrices)~\cite{mieghem2010}.

The leading eigenvalue (i.e., the eigenvalue with the largest magnitude) $\lambda$ of $A$ determines fundamental properties of many dynamical processes on graphs~\cite{mieghem2010,masuda2017,newman2018}. For example, under certain assumptions, the critical coupling strength for the transition to synchrony in the Kuramoto model \cite{restrepo2005} of coupled oscillators and in networks of other coupled dynamical systems~\cite{restrepo2006synchronization} is proportional to $\lambda$. Additionally, $1/\lambda$ gives an estimate of an epidemic threshold for many compartmental models of disease spread on graphs~\cite{yangwang2003}. Relatedly, the percolation threshold for the appearance of a giant component on a graph also involves $1/\lambda$ \cite{bollobas2010}.

In light of the above connections to dynamics, it is relevant to characterize the importances of a graph's nodes and edges using spectral properties of $A$. Determining the importances (i.e., centralities) of network nodes, edges, and other subgraphs is relevant for ranking and other applications~\cite{newman2018}, such as removing specific nodes and edges to contain the spread of an infectious disease \cite{matamalas2018}, reach a percolation threshold~\cite{gao2021}, or minimize congestion in a queueing network~\cite{ying2023}. One can express many centrality measures in terms of the leading eigenvector $v$ of $A$. The prototypical example of such an eigenvector-based centrality is eigenvector centrality \cite{bonacich1972}. Other eigenvector-based centralities include PageRank \cite{brin1998} and hub and authority scores \cite{kleinberg1999}, and generalizations of such centrality measures in multilayer and temporal networks \cite{taylor2017,wu2019,taylor2021}.

In the present paper, we examine \textit{dynamical importance} \cite{restrepo2005}, which is a centrality measure that gives an estimate of the change in $\lambda$ due to network perturbations. Using dynamical importance, we compare the effects of edge removals and edge additions (i.e., two different types of network perturbations~\footnote{Some authors pointedly refer to such as perturbations as ``modifications'' because these changes are of finite size, rather than asymptotically small, for graphs of finite size \cite{liu2003,paton2017}.}) on $\lambda$ for several families of graphs. We then examine network perturbations for the Kuramoto model (which is a system of coupled phase oscillators) on graphs \cite{rodrigues2016}. Under certain assumptions, the Kuramoto model's critical coupling strength, which determines when coupled phase oscillators start to synchronize, is inversely proportional to $\lambda$ \cite{restrepo2005}.

Our paper proceeds as follows. In Sec.~\ref{sec:importance}, we review dynamical importance, compare it to the true change in $\lambda$ for network perturbations, and discuss its accuracy. In Sec.~\ref{sec:evec}, we examine dynamical importance from the perspective of the corresponding change in the leading eigenvector of a graph's adjacent matrix. In Sec.~\ref{sec:kuramoto}, we use dynamical importance to study the effects of network perturbations on a previously derived expression for the order parameter of the Kuramoto model on graphs with approximately-homogeneous degree distributions. Finally, in Sec.~\ref{sec:conclusion}, we conclude and discuss future research directions. In the appendices, we provide a few additional details about some assumptions and other relevant considerations.

In our code repository (see \url{https://github.com/ethanjyoung/dynamical_importance.git}), we provide software to compute dynamical importance and iteratively add edges with the largest dynamical importance to a graph.


\section{Dynamical importance}\label{sec:importance}

We begin by reviewing dynamical importance. To aid our presentation, we summarize our key mathematical notation in Table~\ref{tab:notation}.

\begin{table}[t]
\caption{\label{tab:notation} Summary of our key mathematical notation.}
    \begin{ruledtabular}
        \begin{tabular}{cc}
            \textrm{Notation}&
            \textrm{Meaning}\\
            \colrule
            $\Delta x$ & true change in $x$\\
            $\delta x$ & first-order approximation of $\Delta x$\\
            $\acute{x}$ & $x + \Delta x$\\
            $\grave{x}$ & $x + \delta x$\\
            $\hat{y}$ & quantity computed using $\Delta x$\\
            $\check{y}$ & quantity computed using $\delta x$\\
        \end{tabular}
    \end{ruledtabular}
\end{table}

\begin{figure*}
\centering
\begin{tikzpicture}    
    \node(fig1){\includegraphics[scale=0.65]{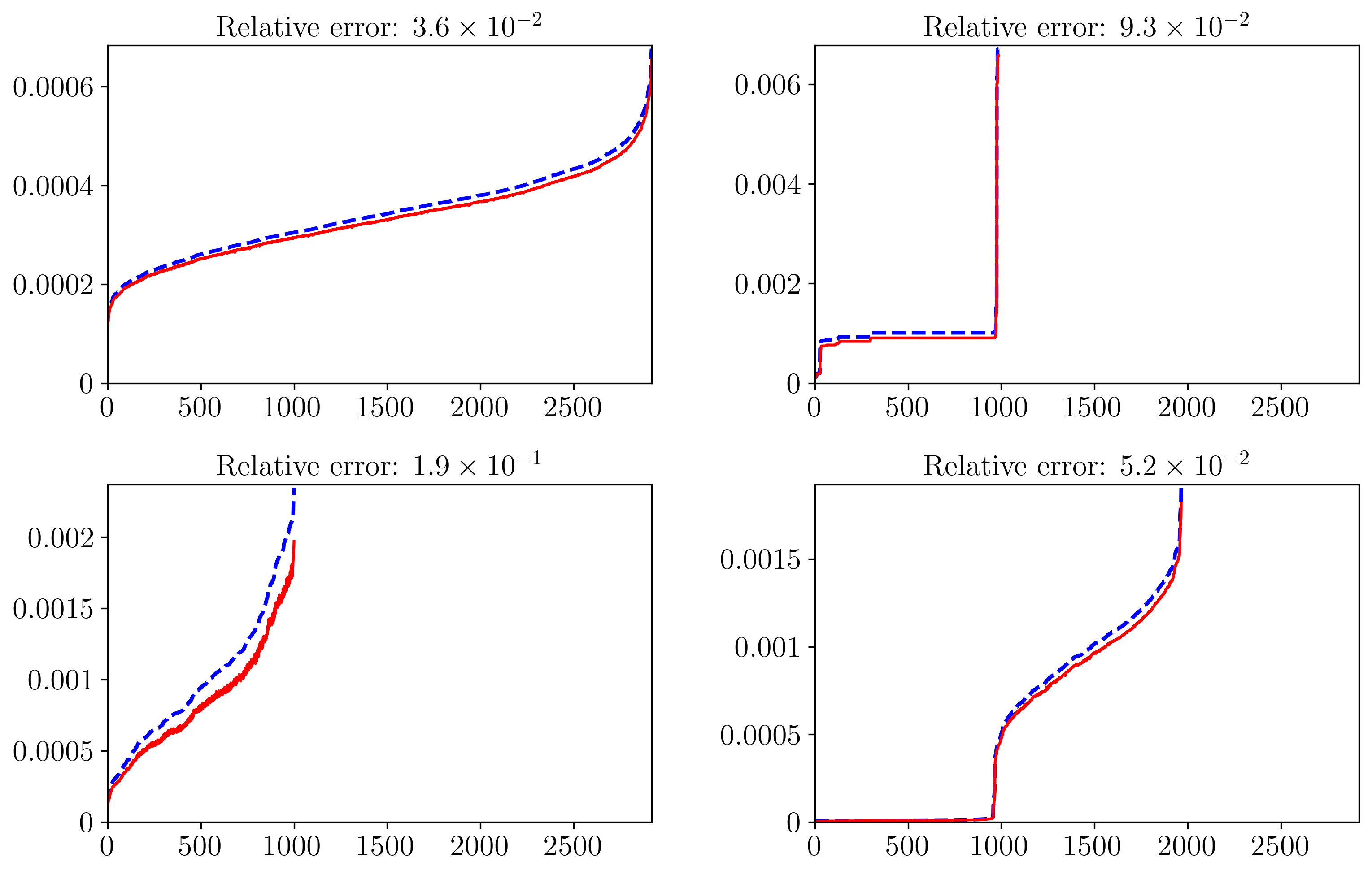}};
    
    \node[below=of fig1, node distance=0cm, yshift=1cm] {edge index (sorted)};
    
    \node[left=of fig1, node distance=0cm, rotate=90,anchor=center,yshift=-0.9cm] {\textcolor{blue}{FoEDI}, \textcolor{red}{$\Delta \lambda / \lambda$}};

 \end{tikzpicture}
 \caption{{We show the first-order edge dynamical importance (FoEDI) (dashed blue curve)  and $\Delta \lambda / \lambda$ (solid red curve) for edge removals in various 200-node graphs. We order the curves for FoEDI and $\Delta\lambda$ by increasing value of FoEDI. The horizontal axis is the edge index. We plot results for graphs that we construct using (top left) the Erd\H{o}s--R\'{e}nyi (ER) model, (top right) the Barab\'asi--Albert (BA) model, (bottom left) the Watts--Strogatz (WS) model, and (bottom right) a stochastic block model (SBM). We describe the parameters of each model in the main text. We use a single instantiation of each type of graph. 
 {In each panel, the number of data points equals the number of edges in the associated graph.} The relative error is $\|x - y\|_2/\|x\|_2$, where $\| \cdot \|_2$ is the $\ell_2$ norm, $x$ is the vector of measured values, and $y$ is the vector of true values.}
}
 \label{fig:edge_remove}
\end{figure*}

\begin{figure*}
\centering 
\begin{tikzpicture}    
    \node(fig2){\includegraphics[scale=0.65]{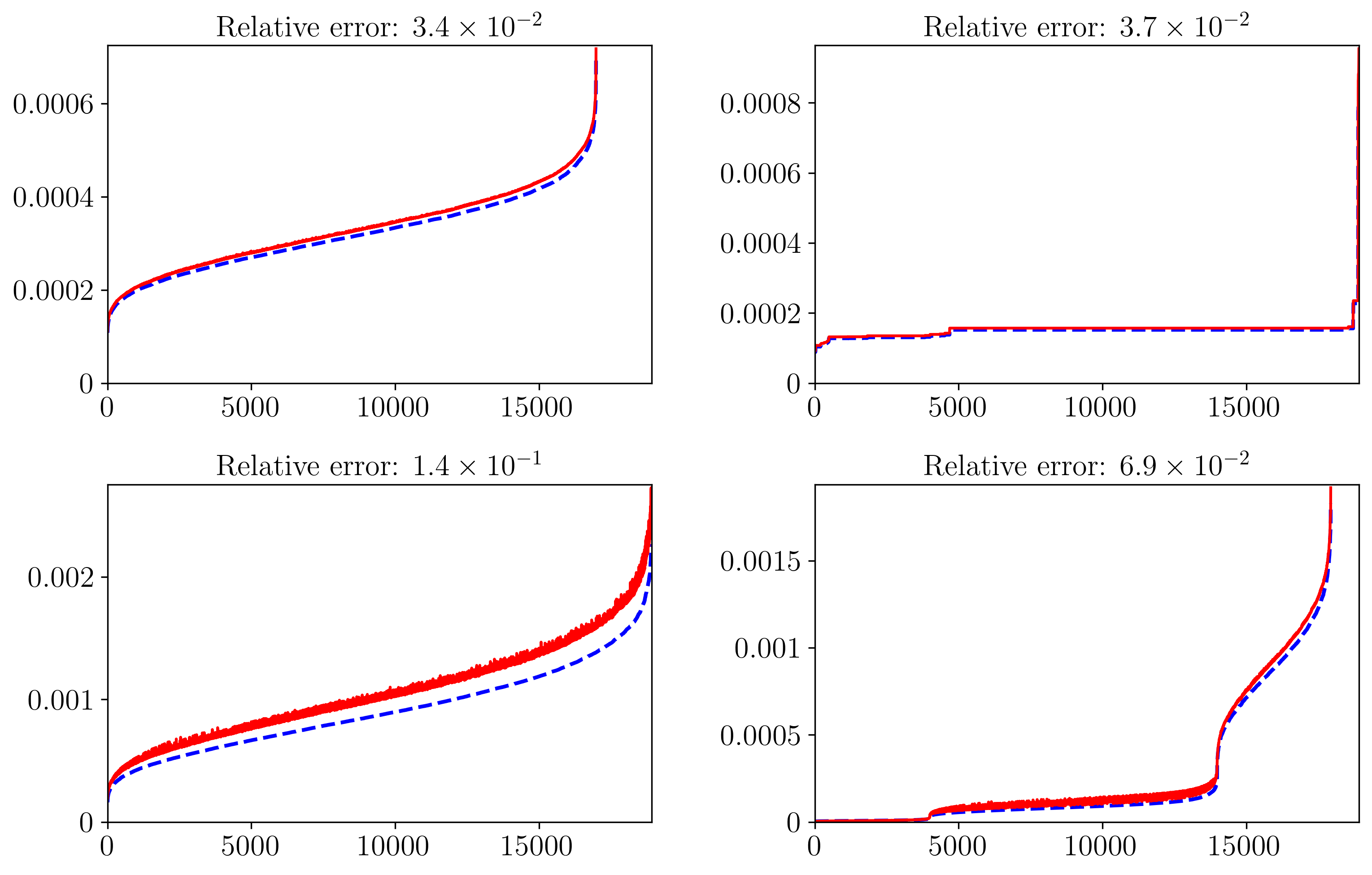}};
    
    \node[below=of fig2, node distance=0cm, yshift=1cm] {edge index (sorted)};
    
    \node[left=of fig2, node distance=0cm, rotate=90,anchor=center,yshift=-0.9cm] {\textcolor{blue}{FoEDI}, \textcolor{red}{$\Delta \lambda / \lambda$}};

 \end{tikzpicture}
\caption{We show the FoEDI (dashed blue curve) and $\Delta \lambda / \lambda$ (solid red curve) for edge additions. We divide $\Delta\lambda$ by $\lambda$ to normalize it. We order FoEDI and $\Delta\lambda$ by increasing FoEDI. We show results for (top left) an ER graph, (top right) a BA graph, (bottom left) a WS graph, and (bottom right) an SBM graph. The random-graph realization in each panel is the same as in the corresponding panel of Fig.~\ref{fig:edge_remove}.}
\label{fig:edge_addition}
\end{figure*}

Given a strongly connected graph (i.e., there is a path from each node to each other node) $G$ with adjacency matrix $A$, leading eigenvalue $\lambda$, leading left eigenvector $u$, and leading right eigenvector $v$, the dynamical importance \cite{restrepo2006dynamical} of the edge $i\rightarrow j$ is
\begin{equation}\label{dynimp}
    \iota_{ij } = \frac{A_{ij}u_iv_j}{\lambda u^Tv}\,.
\end{equation}
Equation \eqref{dynimp} arises from the removal or addition of a single edge.

By the Perron--Frobenius theorem for nonnegative matrices \cite{meyer2023}, the leading eigenvalue $\lambda$ is real and positive, the entries of $u$ all have the same sign, and the entries of $v$ all have the same sign. Without loss of generality, we take the entries of $u$ and $v$ to be nonnegative. We also assume that (1) the graph $G$ is strongly connected and (2) the graph perturbation has a small effect on $\lambda$ and its associated eigenvectors for graphs with $N \gg 1$ nodes.

We start by deriving a first-order approximation of the change $[\Delta\lambda]_{ij}$ in $\lambda$ after the removal or addition of the edge $i \rightarrow j$. The eigenvalue equation for the left and right eigenvectors is
\begin{equation}
    u^T \! Av = \lambda u^T  v\,.
\end{equation}
We use perturbation theory to approximate the change in $\lambda$ for edge removals. Let $A + \Delta A$ denote the new adjacency matrix after the edge removal. Let $\lambda + \Delta \lambda$ denote the associated change in $\lambda$, let $u + \Delta u$ denote the change in $u$, and let $v + \Delta v$ denote the change in $v$. The eigenvalue equation for the perturbed system is
\begin{widetext}
\begin{equation} \label{perturbed}
    (u + \Delta u)^T(A + \Delta A)(v + \Delta v) = (\lambda + \Delta \lambda)(u + \Delta u)^T(v + \Delta v)\,.
\end{equation}
\end{widetext}
We expand \eqref{perturbed} and ignore higher-order terms (i.e., terms that are cubic or have higher powers) 
to obtain
\begin{equation}
    u^T\Delta A \, v = \Delta\lambda \, u^Tv + \text{higher-order terms} \,.
\end{equation}
We divide the retained terms by $u^Tv$ to isolate $\Delta\lambda$ and obtain
\begin{equation}
    \Delta\lambda=\frac{u^T\Delta A \, v}{u^Tv}\,.
\end{equation}
For the removal of the edge $i\rightarrow j$, we have 
\begin{equation}
    [\Delta\lambda]_{ij} = \frac{u_i[\Delta A]_{ij}v_j}{u^Tv}\,.
\end{equation}
Because $[\Delta A]_{ij} = -A_{ij}$, it follows that
\begin{equation}\label{deltalambda}
    [\Delta\lambda]_{ij} = \frac{-A_{ij}u_iv_j}{u^Tv}\,.
\end{equation}

We now derive the dynamical importance $\iota_{ij}$ of the edge $i \rightarrow j$ for edge removals. Consider the relative eigenvalue change
\begin{equation} \label{iota}
    \iota_{ij} = \frac{-[\Delta\lambda]_{ij}}{\lambda}\,,
\end{equation}
where $[\Delta\lambda]_{ij}$ is defined in~\eqref{deltalambda}, we have normalized by the leading eigenvalue $\lambda$, and the factor $-1$ ensures nonnegativity. For edge additions, we do not have the factor $-1$. Inserting \eqref{deltalambda} into \eqref{iota} yields
\begin{equation}
    \iota_{ij} = \frac{A_{ij}u_iv_j}{\lambda u^Tv}\,.
\end{equation}

Henceforth, we only consider graphs that are undirected and unweighted. For undirected graphs, the subscript $ij$ denotes the bidirectional edge $i \leftrightarrow j$, which necessarily includes both $i\rightarrow j$ and $j\rightarrow i$. There are two associated $1$ entries in the perturbation matrix $\Delta A$. We thus have
\begin{equation}
    \iota_{ij} = \frac{u_iv_j+u_jv_i}{\lambda u^Tv}\,.
\end{equation}
The symmetry of $A$ implies that $A$ and $A^T$ have the same eigenvalues, so $u = v^T$. Therefore,
\begin{equation}\label{foedi}
    \iota_{ij} = \frac{2v_iv_j}{\lambda v^Tv}\,.
\end{equation}
We refer to $\iota_{ij}$ as the ``first-order edge dynamical importance'' (FoEDI). The eigenvector $v$ has strictly positive entries, so $\iota_{ij} > 0$.

\begin{algorithm}
    \DontPrintSemicolon
    \caption{Adding edges that maximize the first-order edge dynamical importance $\iota_{ij}$.}
    \KwIn{$E^c$: the set of edges of the complement graph $G^C$}
    \KwOut{$G_{\mathrm{complete}}$: a complete graph}
    \smallskip
    
    $\iota_{G} \gets \{\}$\;

    \For{$k=1$ $\mathbf{to}$ $|E^c|$}{
        \For{$e \in E^c$}{
            Compute $\lambda, v$\;
            Get indices $i$, $j$ of $e$\;
            Compute $\iota_{e} = \frac{2v_iA_{ij}v_j}{\lambda v^Tv}$\;
            $\iota_{G} \gets \iota_{e}$\;
        }
        Get indices $i$, $j$ of max$(\iota_{G})$\;
        Remove edge $e_{ij}$ from $E^c$\;
        Add edge $e_{ij}$ to $G$\;
        $\iota_{G} \gets \{\}$\;
    }
    $G_{\mathrm{complete}} \gets G$\;
    
    \smallskip
    
    \Return{$G_{\mathrm{complete}}$}
    \label{alg:simple}
\end{algorithm}

One can also use \eqref{foedi} to help select which edge (for edge removals) or non-edge (for edge additions) most increases or most decreases $\lambda$. We select one edge at a time, and we use the following procedure (see Algorithm \ref{alg:simple}) to select edges to add to maximize FoEDI. (We use an analogous procedure to select edges to remove to maximize FoEDI.) Given a graph $G$ with associated adjacency matrix $A$, the complement graph $G^C$ is a simple \footnote{A ``simple'' graph is an unweighted, undirected graph with no self-edges or multi-edges~\cite{newman2018}.} graph that consists of all of the edges (except for self-edges) that are not in $G$. The adjacency matrix $A^C$ of $G^C$ is the complement of $A$ and swaps the $0$ entries and $1$ entries of $A$ (except for still having $0$ values on the diagonal). We compute $\iota_{ij}$ for each non-edge of $G$ (i.e., for each edge of $G^C$) and return the non-edge with the edge index that is associated with the largest $\iota_{ij}$. 

Using FoEDI also recovers the Rayleigh quotient $v^T \! Av/v^Tv = \lambda$. See \hyperref[appendix-ray]{Appendix A} for the derivation. This presents opportunities to connect dynamical importance with eigenvalue perturbation theory~\cite{horn2012} (e.g., eigenvalue elasticity). However, many techniques from eigenvalue perturbation theory consider very small perturbations (e.g., changing the weight of an edge in a weighted network), rather than perturbations that change a $1$ into a $0$ (or vice versa) in one entry of an unweighted adjacency matrix $A$~\cite{liu2003, paton2017}.


\subsection{Comparison to the eigenvalue change $\Delta\lambda$}\label{ssec:compare}

In Figs.~\ref{fig:edge_remove} and~\ref{fig:edge_addition}, we compare FoEDI to $\Delta\lambda$ for several graphs. We consider single instantiations of four different types of graphs, which are discussed in detail in the textbook \cite{newman2018}. The first graph is a $G(200,0.15)$ Erd\H{o}s--R\'{e}nyi (ER) graph. The second graph is a Barab\'asi--Albert (BA) graph that we initialize using a single isolated node. The preferential-attachment power is $4$, and we add $5$ edges at each discrete time step. The third graph is a Watts--Strogatz (WS) graph in which each node is initially adjacent to its $k = 4$ nearest neighbors and the rewiring probability is $0.05$. The fourth graph is a stochastic-block-model (SBM) graph with $G(100, 0.2)$ and $G(100, 0.2)$ ER blocks and an independent $0.01$ probability of each edge between nodes in different blocks. 

\begin{figure*}
\centering
\begin{tikzpicture}    
    \node(figb1){\includegraphics[scale=0.65]{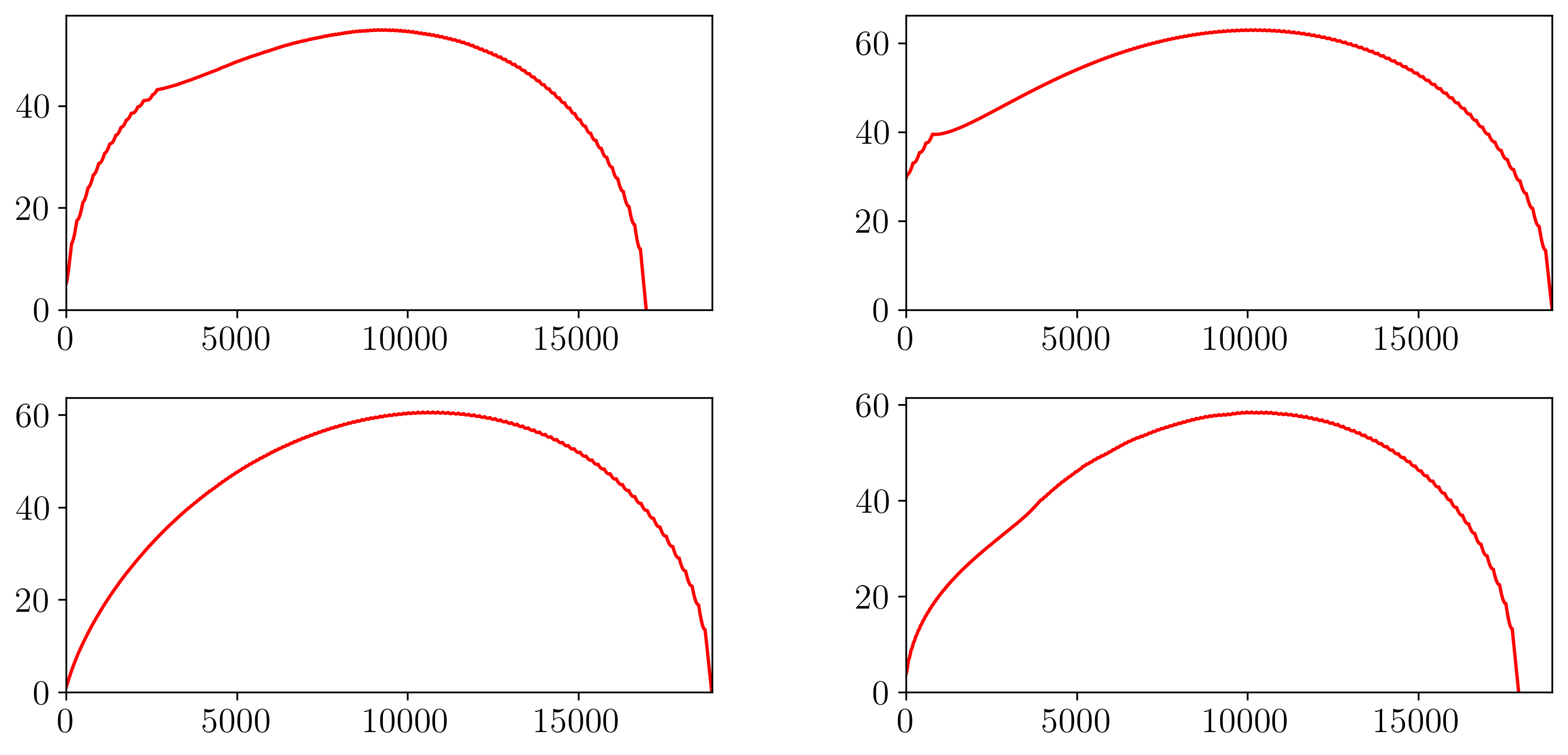}};
    \node[below=of figb1, node distance=0cm, yshift=1cm] {number of edges added};
    \node[left=of figb1, node distance=0cm, rotate=90,anchor=center,yshift=-0.9cm] {$\sigma_d$};
 \end{tikzpicture}
\caption{The standard deviation $\sigma_d$ of the degree distribution of a graph $G$ as we add edges $e_{ij}$, one by one, that maximize $\iota_{ij}$. The horizontal axis is the number of edges that we add from the complement graph $G^C$. We show results for (top left) an ER graph, (top right) a BA graph, (bottom left) a WS graph, and (bottom right) an SBM graph. The random-graph realization in each panel is the same as in the corresponding panel of Fig.~\ref{fig:edge_remove}.}
\label{fig:sd}
\end{figure*}

We show our results for edge removals in Fig.~\ref{fig:edge_remove} and our results for edge additions in Fig.~\ref{fig:edge_addition}. We see that FoEDI always overestimates $\Delta\lambda$ for edge removals and always underestimates $\Delta\lambda$ for edge additions. The difference between FoEDI and the eigenvalue change $\Delta\lambda$ decreases as we increase the numbers of nodes and edges of a graph. In our experiments, this difference is typically negligible. However, this difference is not negligible for additions and removals of nodes. For node dynamical importance, the differences between $\Delta\lambda$ and approximations of it using dynamical importance are significant enough to warrant refining the approximation (e.g., by including second-order terms) to improve its accuracy~\cite{milanese2010, hultgren2011}.
 
As we illustrate in Figs.~\ref{fig:edge_remove} and \ref{fig:edge_addition}, FoEDI estimates $\Delta\lambda$ very accurately for some network models (e.g., ER, BA, and SBM graphs), so higher-order approximations will not meaningfully improve accuracy over $\iota_{ij}$. By contrast, FoEDI is noticeably inaccurate in its estimate of $\Delta\lambda$ for our WS graph. Nevertheless, computing FoEDI still provides a helpful estimate even for this example.

We outline a simple, though computationally expensive, edge-addition procedure using FoEDI. In Fig.~\ref{fig:sd}, we plot the standard deviation $\sigma_d$ of the graph degree distribution as a function of the number of edges that we add using Alg.~\ref{alg:simple}. In Fig.~\ref{fig:eig} in \hyperref[appendix-lambda]{Appendix B}, we plot the change in $\lambda$ as we add edges to each graph. The roughly semicircular curve of $\sigma_d$ of each graph suggests that initially nodes with large degree tend to accrue edges before other nodes. Subsequently, there is not a clear trend in which nodes {have edges added to them}. Additionally, it is well-known that $\lambda \leq d_{\mathrm{max}}$, where $d_{\mathrm{max}}$ is the maximum degree of a graph \cite{mieghem2010}.


\section{Estimating the eigenvector change $\Delta v$}\label{sec:evec}

We now derive an approximation of the change $\Delta v$ in the leading eigenvector of $G$. Consider the expression for FoEDI in \eqref{foedi}, and recall that $u = v^T$ for an undirected graph. The denominator in \eqref{foedi} is constant, so the FoEDI of each edge $e_{ij}$ is determined by the product $v_iv_j$ of the eigenvector entries. Motivated by this product, we examine how much $v$ differs from $v + \Delta v$ by estimating $\Delta v$ for a graph perturbation.

To approximate $\Delta v$, we use the same technique as in our derivation of FoEDI. We begin with the eigenvalue equation
\begin{equation}\label{eval}
    Av = \lambda v\,.
\end{equation}
We perturb \eqref{eval} and write
\begin{equation} \label{eval2}
    (A + \Delta A)(v + \Delta v) = (\lambda + \Delta\lambda)(v + \Delta v)\,.
\end{equation}
Expanding \eqref{eval2} and ignoring terms of second and higher order yields
\begin{equation}
    Av + A\,\Delta v + \Delta A\,v = \lambda v + \lambda\,\Delta v + \Delta \lambda \,v\,,
\end{equation}
which we simplify to obtain
\begin{equation}
    \Delta A \, v - \Delta \lambda \, v = \lambda\,\Delta v - A\,\Delta v\,.
\end{equation}
Therefore,
\begin{equation}
    (\Delta A - \Delta\lambda \,I)v = (\lambda I - A) \Delta v\,,
\end{equation}
where $I$ is the $N \times N$ identity matrix. We cannot compute the inverse of $\lambda I - A$ because it is singular. Therefore, we instead compute the Moore--Penrose generalized inverse \cite{penrose1955}. Let $\acute D = \Delta A - \Delta\lambda \, I$, and let $D = \lambda I - A$. To compute $\acute D$, we use the approximation
\begin{equation}\label{unnorm}
    \iota^\dag_{ij} = \frac{2v_iv_j}{v^Tv}
\end{equation}
of $\Delta\lambda$. The quantity $\iota^\dag_{ij}$ is the unnormalized FoEDI.

\begin{figure*}
\centering
\begin{tikzpicture}    
    \node(figb1){\includegraphics[scale=0.65]{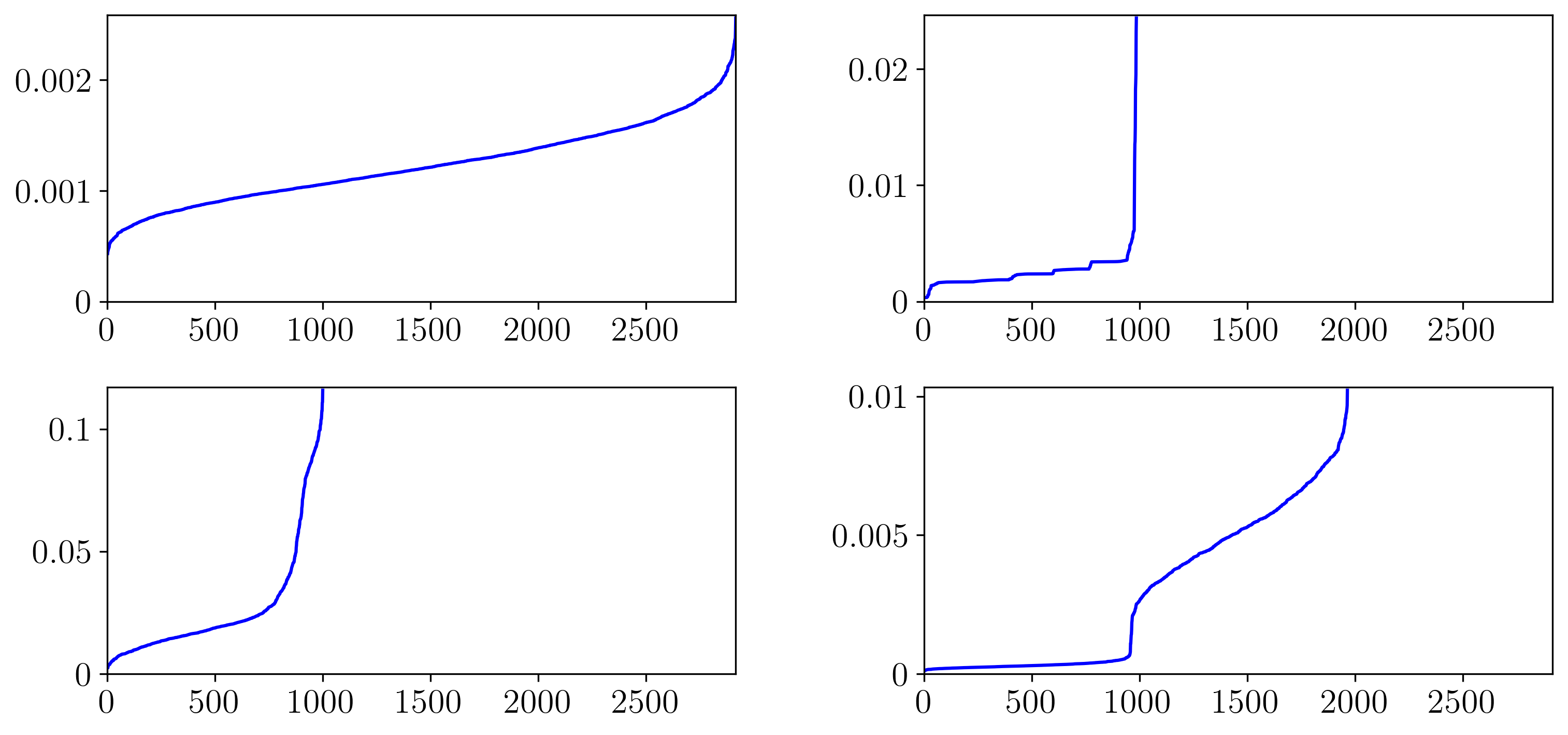}};
    \node[below=of figb1, node distance=0cm, yshift=1cm] {edge index (sorted)};
    \node[left=of figb1, node distance=0cm, rotate=90,anchor=center,yshift=-0.9cm] {relative error};
 \end{tikzpicture}
\caption{The relative error between $\delta v$ and $\Delta v$ for edge removals. The horizontal axis is the index of the edge, and the vertical axis is the relative error. {We order the edge indices by increasing value of the relative error.} We show results for (top left) an ER graph, (top right) a BA graph, (bottom left) a WS graph, and (bottom right) an SBM graph. The random-graph realization in each panel is the same in the corresponding panel of Fig.~\ref{fig:edge_remove}.}
\label{fig:edge_evec_remove}
\end{figure*}

\begin{figure*}
\centering
\begin{tikzpicture}    
    \node(figb1){\includegraphics[scale=0.65]{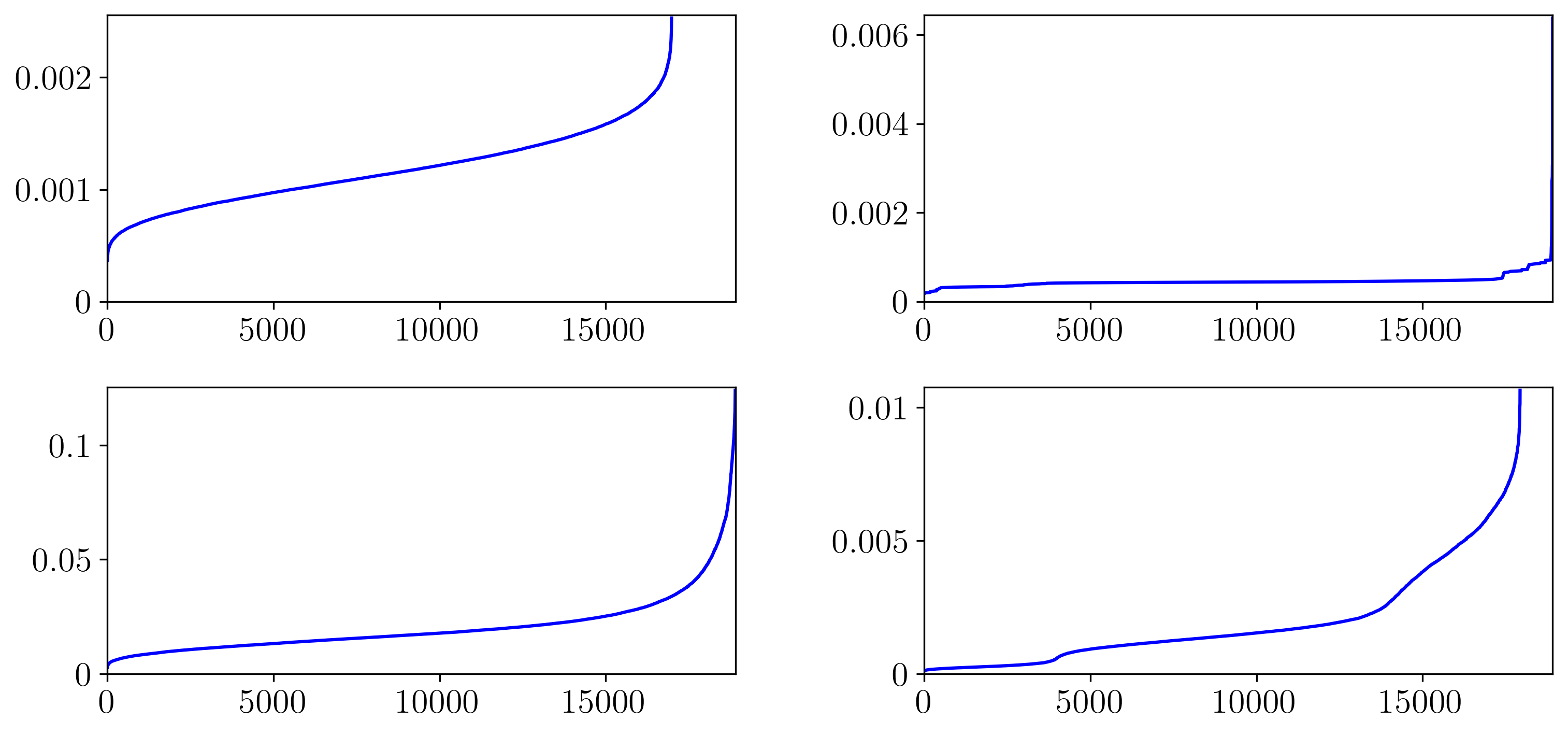}};
    \node[below=of figb1, node distance=0cm, yshift=1cm] {edge index (sorted)};
    \node[left=of figb1, node distance=0cm, rotate=90,anchor=center,yshift=-0.9cm] {relative error};
 \end{tikzpicture} 
\caption{The relative error between $\delta v$ and $\Delta v$ for edge additions. The horizontal axis is the index of the edge, and the vertical axis is the relative error. 
{We order the edge indices by increasing value of the relative error.} We show results for (top left) an ER graph, (top right) a BA graph, (bottom left) a WS graph, and (bottom right) an SBM graph. The random-graph realization in each panel is the same as in the corresponding panel of Fig.~\ref{fig:edge_remove}.}
\label{fig:edge_evec_addition}
\end{figure*}

We seek a first-order approximation $\delta v$ of $\Delta v$. We have 
\begin{equation}
    (\lambda I - A) \delta v = (\Delta A - \Delta\lambda \, I)v\,.
\end{equation}
Solving for $\delta v$ yields
\begin{equation}\label{evec}
    \delta v = D^G\acute Dv\,,
\end{equation}
where $D^G$ denotes the Moore--Penrose generalized inverse of $D$. In Figs.~\ref{fig:edge_evec_remove} and~\ref{fig:edge_evec_addition}, we plot the relative error between the approximation $\delta v$ and the true change $\Delta v$ for the graphs from Fig.~\ref{fig:edge_remove}. We show our results for edge removals in Fig.~\ref{fig:edge_evec_remove} and our results for edge additions in Fig.~\ref{fig:edge_evec_addition}. In both figures, the relative error is particularly large for the WS graph.

Computing $\delta v$ using Eq.~(\ref{evec}) is computationally expensive~\footnote{For an $N \times N$ matrix, computing the Moore--Penrose generalized inverse using the singular-value decomposition (SVD) has complexity $O(N^3)$. However, there exist algorithmic improvements (e.g., that exploit certain matrix structures) that reduce the cost of computing Moore--Penrose inverses. For example, the algorithm in \cite{courrieu2005} that is based on the full-rank Cholesky factorization has a lower-bound complexity of $O(\log N)$ for symmetric, positive-definite matrices.}, but this expression is useful {because we can determine $\delta v$ for any graph perturbation. We can also use $\delta v$ to approximate FoEDI after a perturbation.} Let $\grave{v}$ denote $v + \delta v$, where we compute $\delta v$ from Eq.~\eqref{evec}. Using Eq.~\eqref{unnorm}, we obtain
\begin{equation}
    \check{\iota}^\dag_{ij} = \frac{2\grave{v}_i\grave{v}_j}{\grave{v}^T\grave{v}}\,.
\end{equation}


\subsection{Upper bound on $\Delta v$}\label{ssec:bound}

We now discuss an upper bound on $\Delta v$. It was derived previously in \cite{hultgren2011}. 

Let $\acute v = v + \Delta v$, and let $\lambda_2$ be the second largest (in magnitude) eigenvalue of $A$. We have the upper bound~\cite{hultgren2011}
\begin{equation}\label{bound1}
    \sin(\theta_{v, \acute v}) \leq \frac{\|\Delta A\|_2}{\lambda - \lambda_2}\,,
\end{equation}
where $\theta_{v, \acute v}$ denotes the angle between $v$ and $\acute v$.

Hultgren~\cite{hultgren2011} noted that the inequality~\eqref{bound1} is valid as long as $A$ is symmetric. Because we consider undirected and unweighted graphs, this requirement is satisfied. In this setting, $\|\Delta A\|_2 = 1$, so we further simplify \eqref{bound1} to obtain
\begin{equation}\label{bound2}
    \sin(\theta_{v, \acute v}) \leq \frac{1}{\lambda - \lambda_2}\,.
\end{equation}

Based on our observations, which supports a statement in \cite{restrepo2006dynamical}, it is often the case that (1) $\lambda - \lambda_2$ is large and (2) it is larger for denser graphs, with $\lambda \rightarrow N - 1$ and $\lambda_2 \rightarrow -1$ as the number of edges $m \rightarrow N(N - 1)/2$. However, the bound does not give useful information if $\lambda$ and $\lambda_2$ have similar magnitudes (i.e., if the spectral gap is small). For example, $\lambda - \lambda_2$ can be less than $1$ for graphs with community structure (e.g., as generated using an SBM) and ring-like graphs (e.g., as generated using the WS model).


\section{The Kuramoto model of coupled oscillators}\label{sec:kuramoto}

The Kuramoto model of coupled phase oscillators is a canonical model to study phenomena such as synchronization on networks~\cite{rodrigues2016}. The structure of an underlying network impacts the dynamics of Kuramoto oscillators in interesting ways. We consider the Kuramoto model with diffusive coupling.

We first define the relevant Kuramoto order parameters. The complex-valued ``local order parameter'' of oscillator $i$ is $r_ie^{\mathrm{i}\psi_i}$, where $\mathrm{i} = \sqrt{-1}$ is the imaginary unit and $r_i \in [0,1]$ and $\psi_i \in [-\pi,\pi)$, respectively, are the amount of synchrony and the mean phase of oscillator $i$ and its neighbors. We use the positive real-valued order parameter $r = \lvert\frac{1}{N}\sum^N_{i = 1}r_ie^{\mathrm{i}\psi_i}\rvert$ to measure the overall amount of synchrony of all oscillators.

Given the phase $\Theta_i(t)$ of oscillator $i$, its natural frequency $\omega_i$, and the coupling strength $k$, the Kuramoto model on a graph is the set of coupled ordinary differential equations
\begin{equation}
	\dot{\Theta}_i = \omega_i + k\sum^N_{j = 1}A_{ij}\sin(\Theta_j - \Theta_i)\,.
\end{equation}
The critical coupling strength $k_c$ signifies the onset of a transition to synchronization \cite{rodrigues2016}. Under specific assumptions (see \hyperref[appendix-kuramoto]{Appendix C}), which include the requirement that $\omega_i$ is statistically independent of $r_i$ and $\psi_i$, it has been shown that $k_c \propto 1/\lambda$ \cite{restrepo2005}. Specifically, under those assumptions, the $N \rightarrow \infty$ asymptotic expression for $k_c$ is
\begin{equation}\label{kc}
    k_c = \frac{2}{\pi\lambda g(0)}\,,
\end{equation}
where $g(\omega)$ is the probability distribution from which we draw the natural frequencies.
Let
\begin{equation}\label{eta}
    \eta = \frac{\langle v\rangle^2\lambda^2}{N\langle d\rangle^2\langle v^4\rangle}
\end{equation}
and
\begin{equation}
    \alpha=\frac{- g''(0)}{8g(0)}\,,
\end{equation}
where $\langle d \rangle$ denotes the mean degree of a graph. The square of the order parameter is
\begin{equation}\label{order}
    r^2 = \left(\frac{\pi^2g(0)^2\eta}{4\alpha}\right)\left(\frac{k}{k_c} - 1\right)\left(\frac{k}{k_c}\right)^{-3}\,.
\end{equation}

Restrepo et al.~\cite{restrepo2005} made several assumptions (see \hyperref[appendix-kuramoto]{Appendix C}) to ensure that Eqs.~\eqref{kc} and \eqref{order} are valid. They noted that graphs with an approximately-homogeneous degree distribution (specifically, graphs for which the mean degree $\langle d\rangle\approx\lambda$) guarantee that Eq.~\eqref{order} holds asymptotically and illustrated numerically that Eq.~\eqref{order} is valid for $k/k_c \lessapprox 1.3$. Accordingly, we work in this setting.


\subsection{Estimating the order parameter $r$ with edge additions}

Because of the presence of $\lambda$ and $v$ in Eqs.~\eqref{kc} and \eqref{eta}, we can study how $r$ changes as we add edges to a graph, provided the degree distribution remains approximately homogeneous. Using our estimates of $\Delta\lambda$ in Eq.~\eqref{unnorm} and $\Delta v$ in \eqref{evec}, we obtain perturbed versions of Eqs.~\eqref{kc} and \eqref{eta} after adding the edge $e_{ij}$ to a graph $G$. These expressions are
\begin{equation}\label{kc-hat}
    \hat{k}_c = \frac{2}{\pi(\lambda + \Delta\lambda) g(0)}
\end{equation}
and
\begin{equation}\label{eta-hat}
    \hat{\eta} = \frac{\langle \acute{v}\rangle^2(\lambda+\Delta\lambda)^2}{N(\langle d\rangle + \Delta \langle d\rangle)^2\langle \acute{v}^4\rangle}\,,
\end{equation}
where $\acute{v} = v + \Delta v$.

\begin{figure}[b]
\begin{tikzpicture}
    \node(fig6){\includegraphics[scale=0.65]{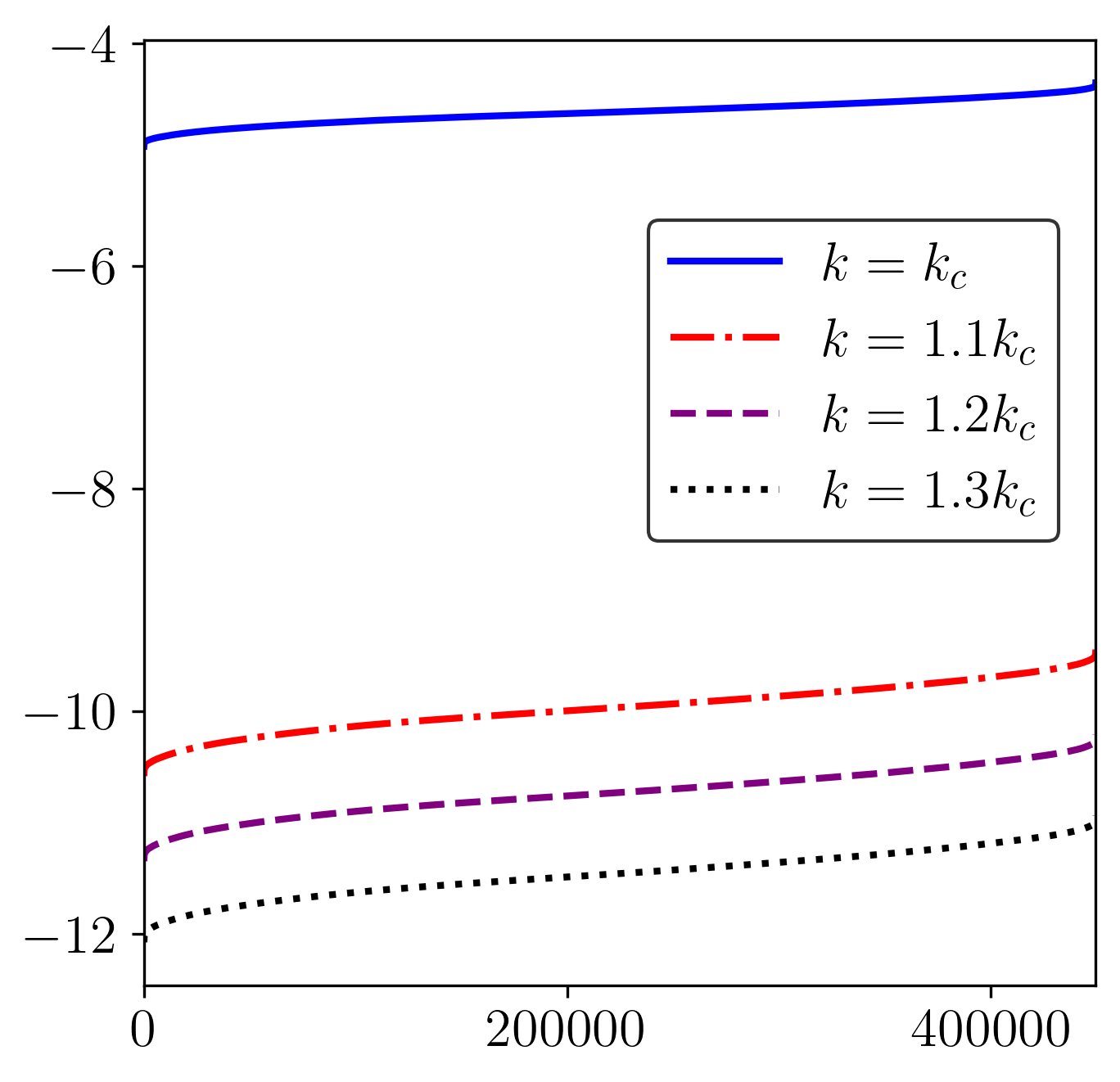}};
    \node[below=of fig6, node distance=0cm, yshift=1cm] {edge index (sorted)};
    \node[left=of fig6, node distance=0cm, rotate=90,anchor=center,yshift=-0.9cm] {ln$(\Delta r)$};
\end{tikzpicture}
\caption{The natural logarithm of the order-parameter change $\Delta r$ for each edge $e_{ij}$ of the complement graph $G^C$ in order of increasing values of $\Delta r$. The horizontal axis is the edge index after we sort the edges. We describe the graph structure and model parameters in the main text. The solid blue curve gives $\Delta r$ for the critical coupling strength $k = k_c$, the dash-dotted red curve gives $\Delta r$ for the coupling strength $k = 1.1k_c$, the dashed purple curve gives $\Delta r$ for $k = 1.2k_c$, and the dotted black curve gives $\Delta r$ for $k = 1.3k_c$.}\label{fig:order-foedi}
\end{figure}

Using Eqs.~\eqref{kc-hat} and \eqref{eta-hat}, the square of the order parameter after adding the edge $e_{ij}$ is
\begin{equation}\label{order-hat}
    \hat{r}^2 = \left(\frac{\pi^2g(0)^2\hat{\eta}}{4\alpha}\right)\left(\frac{k}{\hat{k}_{c}} - 1\right)\left(\frac{k}{\hat{k}_{c}}\right)^{-3}\,.
\end{equation}
Therefore, for a fixed coupling strength $k$, we see that $\hat{r}^2 > r^2$ for edge additions. An analogous derivation gives $\hat{r}^2 < r^2$ for edge removals.

We illustrate Eq.~\eqref{order-hat} with a particular scenario. Consider a graph with $N \gg 1$ nodes such that the degree $d_i$ of each node $i$ satisfies $N \geq d_i \gg 1$ and the mean degree satisfies $\langle d \rangle \approx \lambda$. When we add an edge, the leading eigenvalue $\lambda$ and the mean degree $\langle d \rangle$ all increase by small amounts. Therefore, $\hat{\eta} \approx \eta$ and $\hat{k}_{c} < k_c$, so $\hat{r}^2 > r^2$ for edge additions.
 
We now relate $r$ to FoEDI. The expression $\pi^2g(0)^2\hat{\eta}/(4\alpha)$ in Eq.~\eqref{order-hat} approximately equals $\pi^2g(0)^2\eta/(4\alpha)$ because $\hat{\eta} \approx \eta$ for a graph with an approximately-homogeneous degree distribution. Let $\beta = \pi^2g(0)^2\eta/(4\alpha)$, which is a constant. Substituting $\beta$ and Eq.~\eqref{kc-hat} into Eq.~\eqref{order-hat} yields
\begin{align}\label{subs}
    \hat{r}^2 &= \beta\left(\frac{k}{\hat{k}_{c}} - 1\right)\left(\frac{k}{\hat{k}_{c}}\right)^{-3} \notag \\
    &= \beta\left(\frac{k\pi(\lambda+\Delta\lambda) g(0)}{2} - 1\right)\left(\frac{k\pi(\lambda + \Delta\lambda) g(0)}{2}\right)^{-3}\,.
\end{align}

Let $\gamma = \pi g(0)/2$, which is also a constant. We use the unnormalized FoEDI $\iota^\dag_{ij}$ as a first-order estimate of $\Delta\lambda$ for each edge $e_{ij}$. Substituting $\gamma$ and $\iota^\dag_{ij}$ into Eq.~\eqref{subs} yields
\begin{equation}\label{order-foedi}
    \check{r}^2 = \beta\,(k\gamma(\lambda+\iota^\dag_{ij}) - 1)\,(\gamma (k\lambda + \iota^\dag_{ij}))^{-3}\,.
\end{equation}
Equation~\eqref{order-foedi} expresses the approximate squared order parameter $\check{r}^2$ after an edge perturbation in terms of the coupling strength $k$ and the unnormalized FoEDI $\iota^\dag_{ij}$. As we discussed in Sec.~\hyperref[ssec:compare]{IIA}, the way that one adds edges to a graph can break the approximate homogeneity of a degree distribution. That is, edge additions can cause the standard deviation of the degree distribution to become too large. We do not know an upper bound on the number of edges that one can add and still preserve approximate degree-distribution homogeneity, although one can obtain an approximate bound numerically on a case-by-case basis. We expect that the maximum standard deviation of the degree distribution that retains degree-distribution homogeneity increases with the graph size (i.e., the number of nodes) $N$.

We give an example that demonstrates how to use Eq.~\eqref{order-foedi}. As in \cite{restrepo2005}, we suppose that the distribution of the natural frequencies is $g(\omega) = (3/4)(1 - \omega^2)$ for $-1 < \omega < 1$ and $g(\omega) = 0$ otherwise. We consider a graph with $N = 1000$ nodes and use a configuration model (which is a type of random-graph model) \cite{fosdick2018} without self-edges or multi-edges. The method that we use to prevent self-edges and multi-edges is described in \cite{viger2005}. We choose the node degrees uniformly at random from the set $\{75, 76, \ldots, 124, 125\}$. For each edge $e_{ij}$ of the complement graph $G^C$, we compute Eq.~\eqref{unnorm} and insert it into \eqref{order-foedi}. We consider coupling strengths of $k = k_c$, $k = 1.1k_c$, $k = 1.2k_c$, and $k = 1.3k_c$; equation \eqref{order} is valid for these values.

In Fig.~\ref{fig:order-foedi}, we plot $\ln (\Delta r) = \ln(\acute{r} - r)$ for the different coupling strengths $k$ for a single instantiation of our configuration-model graph. For this graph, $\lambda \approx \langle d \rangle$, so its degree distribution is approximately homogeneous. We observe that $\Delta r$ is smaller for larger multiples of the critical coupling strength $k_c$. This observation is expected, because the order parameter $r$ for the critical coupling strength $k = k_c$ (i.e., when the oscillators begin to synchronize) increases more due to an edge addition than when $k > k_c$. For progressively larger values of $k$, an individual edge addition has a progressively smaller impact on the order parameter. For $k = 1.3k_c$ in our example, adding edges uniformly at random is comparably effective as adding edges that maximize FoEDI.


\section{Conclusions and Discussion}\label{sec:conclusion}

We studied a previously derived measure of edge importance called dynamical importance, which is a first-order approximation of how much the leading eigenvalue $\lambda$ of a graph's adjacency matrix $A$ changes when one removes an edge from it or adds an edge to it \footnote{One can also define a notion of dynamical importance that is based on node perturbations~\cite{restrepo2006dynamical}.}. We examined first-order edge dynamical importance (FoEDI) for undirected, unweighted graphs. We investigated several computational aspects of FoEDI and related it to diffusive Kuramoto dynamics on graphs. We compared FoEDI to the true change in $\lambda$, {derived an approximation of the change in the leading eigenvector $v$ after a graph perturbation}, and expressed the Kuramoto order parameter in terms of FoEDI.

We compared FoEDI to the true change $\Delta\lambda$ in $\lambda$ for Erd\H{o}s--R\'{e}nyi (ER) graphs, Barab\'{a}si--Albert (BA) graphs, Watts--Strogatz (WS) graphs, and stochastic-block-model (SBM) graphs. In our computations, we observed larger relative errors for our WS graph than for the other graphs. We also constructed an edge-addition scheme using FoEDI and observed that large-degree nodes tend to accrue edges earlier than other nodes.

We derived a first-order estimate $\delta v$ of the true change $\Delta v$ of an adjacency matrix's leading eigenvector $v$ after a graph perturbation. Our estimate is computationally infeasible for large graphs, as it involves computing a generalized inverse of a matrix. Nevertheless, its small relative error for some graphs, such as those that are generated by the ER and BA random-graph models, demonstrates its potential usefulness. We also discussed a previously derived upper bound on the angle between $v$ and $v + \Delta v$ for any graph perturbation. We observed that this bound does not give useful information for graphs with small spectral gaps. Interesting future directions to explore include (1) examining the numerical stability of computing generalized inverses of relevant adjacency matrices and (2) exploring how to exploit symmetries and other structures of adjacency matrices to yield better bounds on $\delta v$. Although removing an edge from $A$ or adding an edge to it corresponds to changing two entries of $A$ for an undirected graph (and to one entry for a directed graph), one cannot use methods like the Sherman--Morrison formula \cite{sherman1950} to update the adjacency-matrix inverse because the leading eigenvalue $\lambda$ also changes. To efficiently compute a generalized inverse $X^G$ of a matrix $X$, it seems useful to explore iterative methods that are numerically stable when $X$ is nonsingular \cite{pan2018, hotelling1943}.

We also estimated the change in the order parameter $r$ (which measures the amount of synchronization) in a Kuramoto coupled-oscillator model after edge additions. We focused on an expression for $r$ for graphs with approximately-homogeneous degree distributions (i.e., when $\lambda$ approximately equals the mean degree). In this setting, we derived an expression for the change $\check{r}$ in $r$ in terms of FoEDI. Studying perturbations of $r$ in this manner is efficient and practical because one only needs to compute the eigendecomposition of $A$. Although computing FoEDI grants flexibility in estimating the change in the order parameter for any edge, adding too many edges can break the approximate degree-distribution homogeneity. Our exploration suggests that it may be useful to study the relationship between network perturbations and synchronization. One relevant research direction is to study the number of edges that one needs to add {(for different edge-addition strategies)} to a network of Kuramoto oscillators to achieve global synchrony \cite{townsend2020, kassabov2021, kassabov2022}.

It is common to study the relationship between network perturbations and dynamical processes on networks in the context of controllability of dynamics on graphs (and on more complicated types of networks) \cite{xiang2019}. We expect that edge-perturbation and node-perturbation schemes that are based on dynamical importance are particularly relevant in situations (e.g., in diffusive dynamics) in which spectral information plays a role in determining the behavior of a dynamical process. An important avenue of research involves exploring how dynamics are affected by directed edges and weighted edges (e.g., see \cite{delabays2019}). For instance, one can examine how the perturbations of edge weights (i.e., increasing or decreasing edge weights, without removing or adding any edges) impact the dynamics of a system. These perturbations are different than our paper's perturbations (which are sometimes called ``modifications'' \cite{liu2003, paton2017} or ``network surgery'' \cite{allen2017} because of their finite size for finite-size networks), as they can be infinitesimal in size.


\begin{acknowledgements}

We thank Alex Arenas, James Gleeson, Desmond Higham, Jim Nagy, Piet Van Mieghem, and two anonymous referees for helpful comments.

\end{acknowledgements}


\appendix


\renewcommand{\thefigure}{A.\arabic{figure}}
\setcounter{figure}{0}


\section{Recovering the Rayleigh quotient with FoEDI}\label{appendix-ray}

\begin{figure*}
\centering
\begin{tikzpicture}    
    \node(figb1){\includegraphics[scale=0.65]{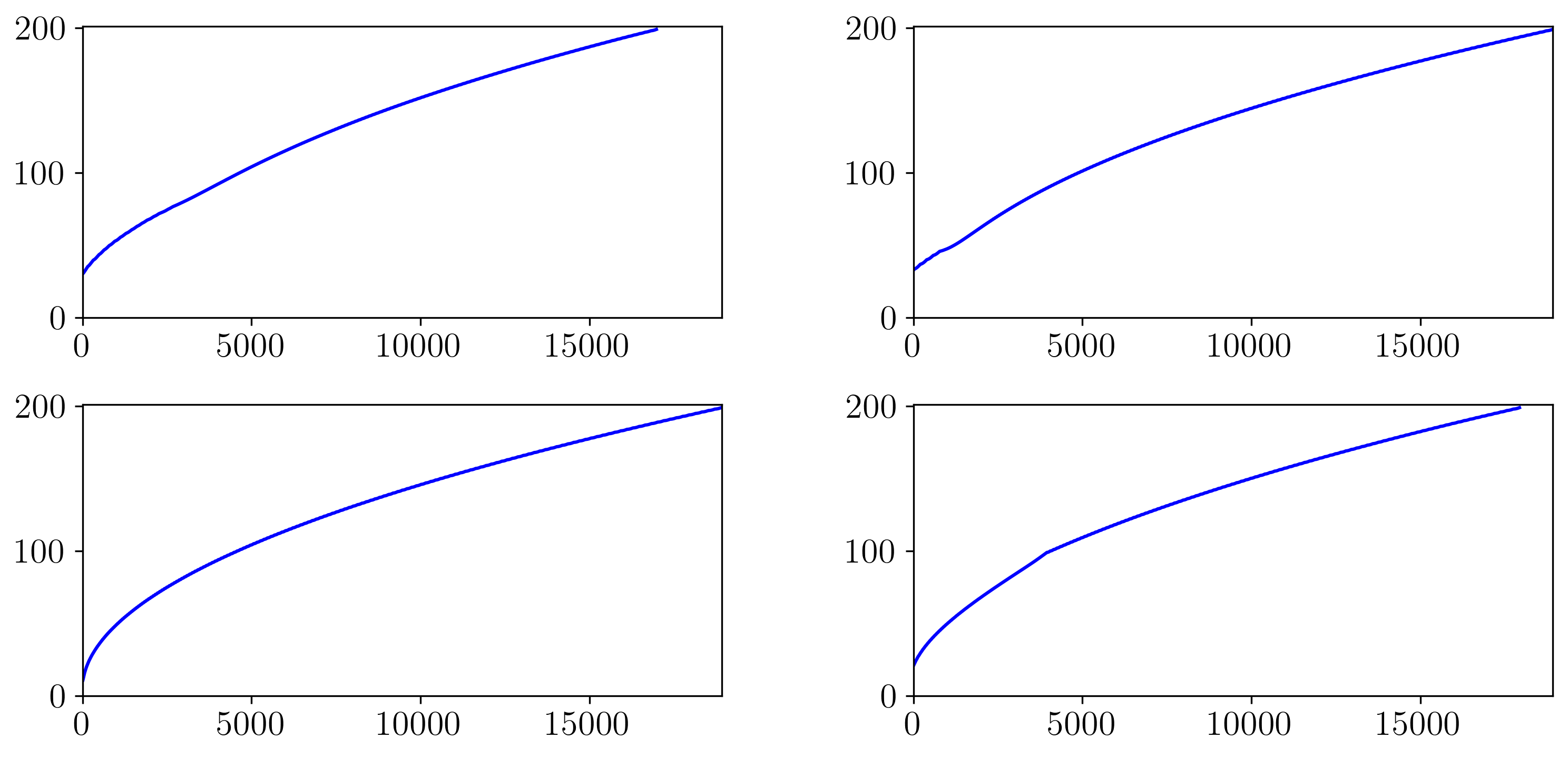}};
    \node[below=of figb1, node distance=0cm, yshift=1cm] {number of edges added};
    \node[left=of figb1, node distance=0cm, rotate=90,anchor=center,yshift=-0.9cm] {$\lambda$};
 \end{tikzpicture}
\caption{The leading eigenvalue $\lambda$ (vertical axis) as we add edges $e_{ij}$, one by one, that maximize the FoEDI $\iota_{ij}$. The horizontal axis is the number of edges that we add from the complement graph $G^C$. We show results for (top left) an ER graph, (top right) a BA graph, (bottom left) a WS graph, and (bottom right) an SBM graph. The random-graph realization in each panel is the same in the corresponding panel of Fig.~\ref{fig:edge_remove}.
}\label{fig:eig}
\end{figure*}

In this appendix, we relate FoEDI to the Rayleigh quotient. Consider an undirected and unweighted graph with adjacency matrix $A$. Expanding Eq.~\eqref{dynimp} yields
\begin{equation}\label{a1}
    \iota_{ij} = \frac{u_iA_{ij}v_j + u_jA_{ji}v_i}{\lambda u^Tv}\,.
\end{equation}
The leading eigenvalue $\lambda$ is a normalization factor in Eq.~\eqref{a1}, so we ignore it and look at the unnormalized FoEDI
\begin{equation}
    \iota^\dag_{ij} = \frac{u_iA_{ij}v_j + u_jA_{ji}v_i}{u^Tv}\,.
\end{equation}
We begin with the sum of $\iota^\dag_{ij}$ over all edges. This sum is
\begin{equation}
    \sum^{N}_{i = 1}\sum^{N}_{j = 1}\iota^\dag_{ij} = \sum^{N}_{i = 1}\sum^{N}_{j = 1}\frac{A_{ij}u_i v_j + A_{ji}u_j v_i}{u^{T}v}\,.
\end{equation}
It follows that
\begin{equation} \label{RHS}
    \sum^{N}_{i = 1}\sum^{N}_{j = 1}\iota^\dag_{ij} = \frac{u^TAv}{u^Tv}\,.
\end{equation}
The right-hand side of \eqref{RHS} is the Rayleigh quotient. Because $A$ is symmetric, $u^T = v$, so we obtain
\begin{equation}
    \sum^{N}_{i = 1}\sum^{N}_{j = 1}\iota^\dag_{ij} = \lambda\,.
\end{equation}


\renewcommand{\thefigure}{B.\arabic{figure}}
\setcounter{figure}{0}


\section{Calculation of the leading eigenvalue $\lambda$ as we add edges}\label{appendix-lambda}

In Fig.~\ref{fig:eig}, we show how $\lambda$ changes as we add edges $e_{ij}$ that maximize $\iota_{ij}$ to a graph $G$. The nonlinear increase in $\lambda$ as a function of the number of added edges is consistent with how $\sigma_d$ (the standard deviation of the degree distribution) changes in Fig.~\ref{fig:sd} as one adds more edges. We expect the slope of $\lambda$ to be large when the slope of $\sigma_d$ is large. Our numerical computations confirm this expectation.


\section{The assumptions on the critical coupling strength $k_c$ and the order parameter $r$}\label{appendix-kuramoto}

We now outline the assumptions that were made in \cite{restrepo2005} to obtain Eqs.~\eqref{kc} and \eqref{order}.

The assumptions to obtain Eq.~\eqref{kc} are as follows: 
\begin{enumerate}[ {(}i{)} ]
  \item the graph $G$ is unweighted and undirected;
  \item the distribution $g(\omega)$ is symmetric about a local maximum (which, without loss of generality, we take to be at $\omega = 0$);
  \item the degree $d_i$ of each node $i \in \{1, \ldots, N\}$ satisfies $d_i \gg 1$; and
  \item the existence of solutions $r_i$ (i.e., the positive real-valued order parameter of oscillator $i$ {and its neighbors}) and $\psi_i$ (i.e., the mean phase of oscillator $i$ {and its neighbors}) are statistically independent of $\omega_i$. \medskip
\end{enumerate}


The assumptions to obtain Eq.~\eqref{order} are as follows: (i) $k \approx k_c$ and (ii) $\langle d^4 \rangle$ is finite.




\providecommand{\noopsort}[1]{}\providecommand{\singleletter}[1]{#1}%
%


\end{document}